\documentclass[aps,pre,superscriptaddress,twocolumn]{revtex4-1}

\usepackage{amsmath}
\usepackage{float}
\usepackage{amsfonts}
\usepackage{graphicx}
\usepackage{xcolor}
\usepackage{epstopdf}

\begin{document}


\title{Effects of spatial and temporal noise on a cubic-autocatalytic reaction-diffusion model}


\author{Jean-S\'{e}bastien Gagnon}
\email[]{gagnon01@fas.harvard.edu}
\affiliation{Department of Earth and Planetary Sciences, Harvard University, Cambridge, Massachusetts, USA}

\author{David Hochberg}
\email[]{hochbergd@cab.inta-csic.es}
\affiliation{Centro de Astrobiolog\'{i}a (CSIC-INTA), Madrid, Spain}

\author{Juan P\'{e}rez-Mercader}
\email[]{jperezmercader@fas.harvard.edu}
\affiliation{Department of Earth and Planetary Sciences, Harvard University, Cambridge, Massachusetts, USA}
\affiliation{Santa Fe Institute, Santa Fe, New Mexico, USA}


\date{\today}

\begin{abstract}
We characterize the influence that external noise, with both spatial
and temporal correlations, has on the scale dependence of the
reaction parameters of a cubic autocatalytic reaction diffusion
(CARD) system. Interpreting the CARD model as a primitive reaction
scheme for a living system, the results indicate that power-law
correlations in environmental fluctuations can either decrease or
increase the rates of nutrient decay and the rate of autocatalysis
(replication) on small spatial and temporal scales.
\end{abstract}


\maketitle


\section{Introduction \label{sec:Introduction}}

Reaction-diffusion equations can be used to model various phenomena
ranging from (biological) pattern
formation~\cite{Turing_1952,Meinhardt_1982,Cross_Hohenberg_1993,Tompkins_etal_2014},
ecological invasions~\cite{Holmes_etal_1994}, tumour
growth~\cite{Lowengrub_etal_2010} and oscillating chemical
reactions~\cite{Cross_Hohenberg_1993,Epstein_Pojman_1998}.  Of
particular interest is the cubic autocatalytic reaction-diffusion
(CARD) model~\cite{Pearson_1993,Lesmes_etal_2003}, based on a two
species autocatalytic chemical
reaction~\cite{Higgins_1964,Selkov_1968,Gray_Scott_1983,Gray_Scott_1984,
Gray_Scott_1985,Prigogine_Nicolis_1977}. Numerical simulations of
the deterministic~\cite{Pearson_1993} and
stochastic~\cite{Lesmes_etal_2003} CARD model show the appearance of
self-replicating domains that are analogous to simple cells. This
makes the CARD model a very interesting nontrivial simplified model
for carrying out both analytic and numerical studies of primitive
analogues of living organisms.

Any living organism is in contact and interacts with an environment
that can affect its evolution in important ways. In chemical systems
as well, there is a strong interplay between the microscopic
dynamics and the environmental fluctuations acting on the smallest
spatial and temporal scales. In order to get a quantitative handle
on this phenomena, application of the renormalization group is
ideally suited. One then obtains flow equations which indicate how
certain model parameters can change with the scale of the
observation (or the scale of the probe employed to such end). Our
aim is to get a better understanding of how correlations in external
noise modifies the parameters appearing in simple chemical reaction
models.

In the CARD model, the effect of the environment can be modeled by
adding a noise term to the evolution equations. In previous
works~\cite{Gagnon_etal_2015,Gagnon_PerezMercader_2015}, we
investigated the small-scale properties of the stochastic CARD model
due to the presence of environmental fluctuations using field
theoretical and renormalization techniques. Some technical aspects,
having to do with regularization procedures, made this analysis more
intricate than would have been naively expected, so we focussed our
investigation on noise with only power-law correlations in space. In
this paper, we extend our analysis to noise with both spatial and
temporal power-law correlations.

The remainder of this paper is organized as follows.  In
Sect.~\ref{sec:CARD_model} we briefly present the CARD model and
some of its properties. We then discuss one-loop corrections to the
parameters of the model and the corresponding $\beta$-function in
Sect.~\ref{sec:Renormalization}. We consider the effect of
fluctuations on the parameters in Sect.~\ref{sec:Discussion}.  We
finally conclude in Sect.~\ref{sec:Conclusion}. Feynman rules and
technical details related to the computation of one-loop corrections
are relegated to the appendices.

\section{The stochastic CARD model \label{sec:CARD_model}}

The CARD model~\cite{Pearson_1993,Lesmes_etal_2003} is based on the following chemical
reactions~\cite{Higgins_1964,Selkov_1968,Gray_Scott_1983,Gray_Scott_1984, Gray_Scott_1985,Prigogine_Nicolis_1977}:
\begin{eqnarray}
\label{eq:Gray_Scott_reactions}
\mbox{U} + 2\mbox{V} & \stackrel{\lambda}{\rightarrow} & 3\mbox{V}, \nonumber \\
\mbox{V} & \stackrel{r_{v}}{\rightarrow} & \mbox{P}, \nonumber \\
\mbox{U} & \stackrel{r_{u}}{\rightarrow} & \mbox{Q}, \nonumber \\
  & \stackrel{f}{\rightarrow} & \mbox{U}.
\end{eqnarray}
A substrate U (viewed as the ``food'' in the living system interpretation of the CARD model) is
fed into the system at a constant rate $f$.  The species V (viewed as the ``organism'') consumes the
substrate U and turns it into V via an autocatalytic reaction with rate constant~$\lambda$.  This autocatalytic r
eaction embodies a crude form of metabolism.  In numerical simulations, the species V forms cell-like domains over
the substrate U in a certain parameter range.  Both species V and U decay into inert products P and Q with decay rates $r_{v}$ and $r_{u}$.

The space-time evolution of the chemical concentrations $U({\bf
x},t)$ and $V({\bf x},t)$ in the general case where diffusion and
noise are present is governed by the following equations:

\begin{eqnarray}
\label{eq:Gray_Scott_equations_1}
\frac{\partial V}{\partial t} & = & D_{v}\nabla^{2}V - r_{v}V + \lambda UV^{2} + \eta_{v}(x), \\
\label{eq:Gray_Scott_equations_2}
\frac{\partial U}{\partial t} & = & D_{u}\nabla^{2}U  - r_{u}U - \lambda UV^{2} + \eta_{u}(x) + f,
\end{eqnarray}
where we use the shortcut notation $x = ({\bf x},t)$, $U = U({\bf
x},t)$, $V = V({\bf x},t)$, and $D_{u}$ and $D_{v}$ are the
diffusion constants of the chemical species U and V, respectively.
The terms $\eta_{u}(x)$ and $\eta_{v}(x)$ are additive spacetime-
dependent noises discussed in more detail below.

For simplicity, we work in an approximation where $U \gg 2f/r_{u}$.
In such a case, the feeding term $f$ in
Eq.~(\ref{eq:Gray_Scott_equations_2}) can be
neglected~\cite{Gagnon_etal_2015}.  This approximation is generally
valid when the initial amount of food in the system is large
compared to the equilibrium value, in which case the effects of
feeding are negligible.  In practice, neglecting $f$ amounts to
neglecting one tadpole diagram at one-loop.  For more discussion on this approximation, we refer the reader to Ref.~\cite{Gagnon_etal_2015}.


The effect of the environment on the CARD dynamics is modeled by
adding spacetime dependent noise t erms as indicated in
Eqs.~(\ref{eq:Gray_Scott_equations_1})-(\ref{eq:Gray_Scott_equations_2}).
Examples relevant to chemistry and biology include thermal
fluctuations due to random motion of molecules in chemical reactions
(e.g.~\cite{MacDonald_2006,Gillespie_2007}), mechanical noise in
chemical reactions~\cite{Carnall_etal_2010} and noisy gene
expression (e.g.~\cite{Tsimring_2014}).  The stochastic CARD model
is studied numerically in
Ref.~\cite{Lesmes_etal_2003,Munteanu_Sole_2006}, where it is shown
that the types of patterns produced (growing stripes, cell-like
domains) depend on the amplitude of the (white) noise. The above is
a good example of noise-controlled pattern selection and is
important for chemical and biological applications.

In the following, we use a three-parameter (an amplitude plus two
power-law exponents) Gaussian noise with both spatial and temporal
power-law correlations~\cite{Hochberg_etal_2003} to describe the
stochastic component in
Eqs.~(\ref{eq:Gray_Scott_equations_1})-(\ref{eq:Gray_Scott_equations_2}).
Its statistical properties are given by:
\begin{eqnarray}
\label{eq:Noise_property_1}
\langle \eta_{v}(k) \rangle & = & \langle \eta_{u}(k) \rangle \;\;\;=\;\;\; 0, \\
\label{eq:Noise_property_2}
\langle \eta_{v}(k)\eta_{v}(p) \rangle & = & 2A_{v} |{\bf k}|^{-y_{v}} \omega^{-2\theta_{v}} (2\pi)^{d_{s}+1}\delta^{(d_{s}+1)}(k+p), \\
\label{eq:Noise_property_3}
\langle \eta_{u}(k)\eta_{u}(p) \rangle & = & 2A_{u}|{\bf k}|^{-y_{u}}\omega^{-2\theta_{u}}(2\pi)^{d_{s}+1}\delta^{(d_{s}+1)}(k+p), \\
\label{eq:Noise_property_4}
\langle \eta_{v}(k)\eta_{u}(p) \rangle & = & \langle \eta_{u}(k)\eta_{v}(p) \rangle \;\;\;=\;\;\; 0,
\end{eqnarray}
where we use the shortcut notation $k = ({\bf k},\omega)$ and we
have expressed the correlations in Fourier space for later
convenience. All higher order moments are zero (Gaussian noise) and
$d_{s}$ is the dimension of space.  The noise amplitudes $A_{u},
A_{v} > 0$ are free parameters of the model and give the overall
strength of the fluctuations.  The spatial noise exponents $y_{u},
y_{v}$ and temporal noise exponents $\theta_{u}, \theta_{v}$ are
also free parameters of the model that give the strength of
correlations as a function of wavenumber and frequency. The case of
pure spatial correlations ($\theta_{u} =0$, $\theta_{v} = 0$) is
treated in Ref.~\cite{Gagnon_etal_2015}.  In this paper, we study
the more general case where both spatial and temporal correlations
can be present and acting together.

Note that the additive noise terms in
Eqs.~(\ref{eq:Gray_Scott_equations_1})-(\ref{eq:Gray_Scott_equations_2})
are examples of extrinsic noise, i.e. noise caused by the
application of a random force external to the
system~\cite{vanKampen_2007}. Intrinsic noise is another generic
type of noise that is present even for systems in complete
isolation. It is generally attributed to the fact that chemical
systems are made of discrete particles and quantum mechanical
randomness~\cite{vanKampen_2007}. On physical grounds, we expect
intrinsic noise to be negligible in macroscopic systems at relevant
experimental temperatures, in the same way that quantum fluctuations
are negligible compared to thermal fluctuations in high-temperature
condensed matter systems. In situations where discretization effects
are not negligible, both types of noise must be taken into account.
This can be done using a Master equation approach
(e.g.~\cite{vanKampen_2007,Tauber_etal_2005,Cooper_etal_2013}),
which lies beyond the scope of this paper.

\section{UV renormalization of the stochastic CARD model \label{sec:Renormalization}}

We are interested in the effects of spatially and temporally
correlated noise on the small-scale properties of the CARD model's
dynamics. To do that, we use the renormalization group and run it
from large to small scales, or from the infrared (IR) to the
ultraviolet (UV). The change in model parameters induced by
fluctuations is encoded in $\beta$-functions, thus our goal is to
compute those $\beta$-functions at one-loop in perturbation theory.
Since this type of computation has been carried out in detail for
spatially correlated noise in Ref.~\cite{Gagnon_etal_2015}, we focus
here on the modifications and differences when including explicit
temporal correlations.

\subsection{One-loop corrections to the parameters \label{sec:One_loop_corrections}}

\begin{figure}
\includegraphics[width=0.35\textwidth]{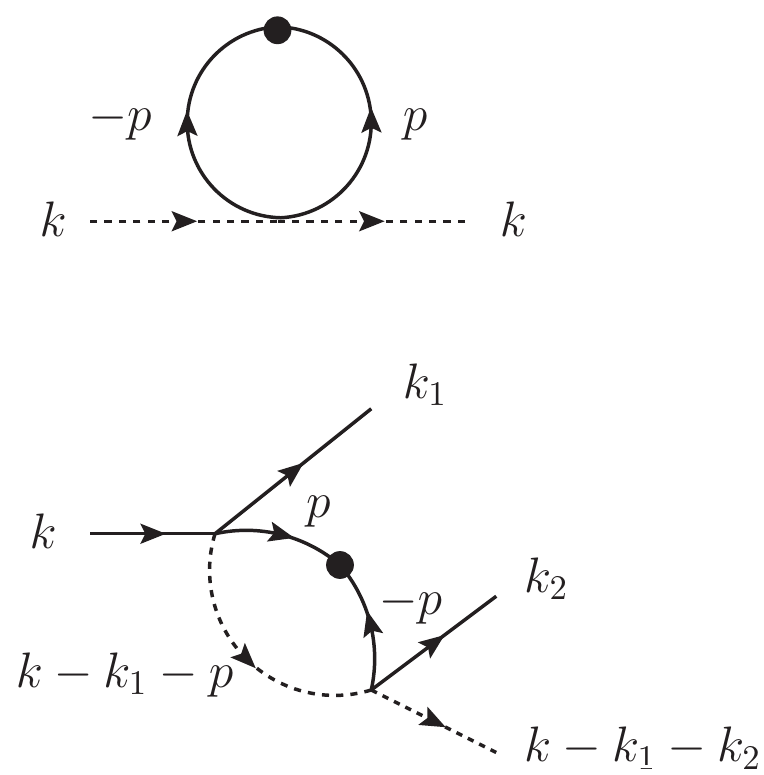}
\caption{One-loop diagrams contributing to $r_{u}$ (top diagram) and
$\lambda$ (bottom diagram), in the approximation $U \gg 2f/r_{u}$.
\label{fig:One_loop_corrections}}
\end{figure}

Beta functions are computed from the UV divergence structure of
Feynman diagrams. The Feynman rules corresponding to the CARD
equations~(\ref{eq:Gray_Scott_equations_1})-(\ref{eq:Gray_Scott_equations_2})
are discussed in Refs.~\cite{Hochberg_etal_2003,Gagnon_etal_2015}
and summarized in Appendix~\ref{sec:Feynman_rules}. At one-loop
order, the only nontrivial corrections to the model parameters are
shown in Fig.~\ref{fig:One_loop_corrections}
\cite{Gagnon_etal_2015}.

To illustrate the differences between the cases with and without temporally correlated noise,
consider the one-loop correction to the decay rate $r_{u}$.
In $d_{s}$ spatial dimensions, the correction is given by:
\begin{eqnarray}
\Gamma_{r_{u}}(0) & = & -2\lambda A_{v} \int\frac{d^{d_{s}}p}{(2\pi)^{d_{s}}} \int\frac{d\omega}{(2\pi)}\; \omega^{-2\theta_{v}} |{\bf p}|^{-y_{v}}\; \nonumber  \\
               &   & \times \left(\frac{1}{D_{v}|{\bf p}|^{2} - i\omega + r_{v}}\right) \left(\frac{1}{D_{v}|{\bf p}|^{2} + i\omega + r_{v}}\right), \nonumber \\
               \label{eq:One_loop_correction_decay_rate_2}
               & = & -2\lambda A_{v} \int\frac{d^{d_{s}}p}{(2\pi)^{d_{s}}} \int\frac{d\omega}{(2\pi)}\; \omega^{-2\theta_{v}} |{\bf p}|^{-y_{v}}\; \nonumber  \\
               &   & \times \left(\frac{1}{\omega^{2} + (D_{v}|{\bf p}|^{2} + r_{v})^{2}}\right),
\end{eqnarray}
where we have taken all external momenta and frequencies to be zero
(this is sufficient for $\beta$-function computations). Contrary to
quantum field theory, there is no Lorentz invariance in the CARD
model, thus the integrals over the frequency and wavenumber in
Eq.~(\ref{eq:One_loop_correction_decay_rate_2}) must be carried out
separately. Depending on the dimension of space $d_{s}$ and on the
parameters of the model $y_{v}$ and $\theta_{v}$,  either the
momentum, or the frequency or even both integrals may diverge.  We
use dimensional continuation as our regulator (see
Ref.~\cite{Gagnon_etal_2015} for a discussion on the use of
dimensional regularization in the presence of power-law noise).
Analytically continuing the space dimension to $d$ and the time
dimension to $z$, we obtain:
\begin{eqnarray}
\label{eq:One_loop_correction_ru_regulated}
\Gamma_{r_{u}}(0) & = & -2\lambda^{(d)} A^{(d,z)}_{v} \int\frac{d^{d}p}{(2\pi)^{d}} \int\frac{d^{z}\omega}{(2\pi)^{z}}\; |\omega|^{-2\theta_{v}} |{\bf p}|^{-y_{v}}\; \nonumber  \\
               &   & \times \left(\frac{1}{|\omega|^{2} + (D_{v}|{\bf p}|^{2} + r_{v})^{2}}\right),
\end{eqnarray}
where $\lambda^{(d)}$ and $A^{(d,z)}_{v}$ are the rate constant and noise amplitude
in $d$ spatial and $z$ time dimensions, respectively.
Simple power counting shows that the one-loop correction behaves as:
\begin{eqnarray}
\Gamma_{r_{u}} & \sim & \Lambda^{d-y_{v}+2z-4\theta_{v}-4},
\end{eqnarray}
where $\Lambda$ is a large momentum scale cutoff.  For convenience,
we define the following parametrization:
\begin{eqnarray}
\label{eq:Definition_1}
\frac{d}{2} - \frac{y_{v}}{2} & = & m - \frac{\epsilon}{2} \hspace{0.8in} m = 1,2,3,\dots \\
\label{eq:Definition_2}
\frac{z}{2} - \theta_{v} & = & \frac{(n+1)}{2} - \frac{\delta}{2} \hspace{0.45in} n = 0,1,2,3,\dots
\end{eqnarray}
with $0 < \epsilon < 2$ and $0 < \delta < 1$, giving:
\begin{eqnarray}
\label{eq:UV_divergences}
\Gamma_{r_{u}} & \sim & \Lambda^{2m + 2n -2 - (\epsilon + 2\delta)}.
\end{eqnarray}
The integers $m$ and $n$ define the order of the divergence
(logarithmic, quadratic, ...) and $\epsilon$ and $\delta$ the
(fractional) distance from some critical dimension
\footnote{Critical dimensions in the present context are defined as
the dimension above which a specific parameter (e.g. $r_{u}$,
$\lambda$) requires renormalization.  For example, corrections to
$r_{u}$ are divergent when $d_{s} \geq d_{c}^{r_{u}} = y_{v} +
4\theta_{v} + 2$ and corrections to $\lambda$ are divergent when
$d_{s} \geq d_{c}^{\lambda} = y_{v} + 4\theta_{v} + 4$ in the CARD
model.}. Note that we are interested in the small-scale properties
of the CARD model and thus in UV divergences, hence the restrictions
on $m$ and $n$ (non-negative values) in
Eqs.~(\ref{eq:Definition_1})-(\ref{eq:Definition_2}).  Note also
that the divergence structure of $\Gamma_{r_{u}}$ depends on $m$ and
$n$, implying that it can be controlled externally via the noise
exponents $y_{v}$ and $\theta_{v}$.

The correction $\Gamma_{r_{u}}$ is logarithmically divergent for $(m
= 1, n = 0)$. This case corresponds to purely spatial power-law
noise and is treated in Ref.~\cite{Gagnon_etal_2015}. Note that it
is not possible to get a logarithmically divergent correction to
$r_{u}$ from purely temporal power-law noise (i.e. $m = 0$, $n =
1$), since $m = 0$ corresponds to IR
divergences~\cite{Gagnon_etal_2015} that are not the focus of the
present study. There are two ways to obtain a quadratically
divergent correction to $r_{u}$.  The first way is by setting ($m =
2$, $n = 0$), corresponding to purely spatial power-law noise and
treated in Ref.~\cite{Gagnon_etal_2015}.  The second way is by
setting ($m = 1$, $n = 1$), and is the first nontrivial case
involving a mixture of spatial and temporal power-law noise. Higher
order divergences are obtained when $m + n > 2$, however their
treatment is more subtle and requires the introduction of higher
order terms in
Eqs.~(\ref{eq:Gray_Scott_equations_1})-(\ref{eq:Gray_Scott_equations_2}),
as discussed in
Refs.~\cite{Gagnon_etal_2015,Gagnon_PerezMercader_2015}.  In the
following, we focus on the ($m=1$, $n=1$) case and leave the
consideration of higher-order divergences for a future work.

Integrating Eq.~(\ref{eq:One_loop_correction_ru_regulated}) over frequency using the method of dimensional regularization
in the presence of noise~\cite{Gagnon_etal_2015}, we obtain:
\begin{eqnarray}
\label{eq:One_loop_correction_ru_regulated_1}
\Gamma_{r_{u}}(0) & = & -\lambda^{(d)} A_{v}^{(d,z)} \;K_{z}\; \Gamma(-\frac{z}{2} + \theta_{v} + 1)\Gamma(\frac{z}{2} - \theta_{v}) \nonumber \\
               &   & \times \int\frac{d^{d}p}{(2\pi)^{d}}\; |{\bf p}|^{-y_{v}} \frac{1}{(D_{v}|{\bf p}|^{2} + r_{v})^{-z+2\theta_{v} + 2}}, \nonumber \\
\end{eqnarray}
where $K_{z} \equiv 2/[(4\pi)^{\frac{z}{2}}\Gamma(\frac{z}{2})]$.
We see from Eq.~(\ref{eq:One_loop_correction_ru_regulated_1}) that the frequency
and momentum integrals mix in a non-trivial way due to the presence of $\theta_{v}$ in
the exponent of $(D_{v}|{\bf p}|^{2} + r_{v})$.  The momentum integral is performed in a similar fashion, giving:
\begin{eqnarray}
\label{eq:One_loop_correction_ru_regulated_2}
\Gamma_{r_{u}}(0) & = & -\frac{\lambda^{(d)} A_{v}^{(d,z)}}{2D_{v}^{-z+2\theta_{v}+2}} K_{d}K_{z} \; \Gamma(-\frac{z}{2} + \theta_{v} + 1)\Gamma(\frac{z}{2} - \theta_{v}) \nonumber \\
               &   & \times \frac{\Gamma\left(-\frac{d}{2}+\frac{y_{v}}{2}-z+2\theta_{v}+2\right)\Gamma\left(\frac{d}{2}-\frac{y_{v}}{2}\right)}{\Gamma\left(-z+2\theta_{v}+2\right)} \nonumber \\
               &   & \times \left(\frac{r_{v}}{D_{v}}\right)^{\frac{d}{2}-\frac{y_{v}}{2}+z-2\theta_{v}-2}.
\end{eqnarray}
We have explicitly checked the commutativity of the two integrals in
Eq.~(\ref{eq:One_loop_correction_ru_regulated}). Substituting the
definitions~(\ref{eq:Definition_1})-(\ref{eq:Definition_2}) (with
$m=1$ and $n=1$) into
Eq.~(\ref{eq:One_loop_correction_ru_regulated_2}) and expanding around
the critical dimension $d_{c}^{\lambda} = y_{v} + 2$ we obtain:
\begin{eqnarray}
\label{eq:One_loop_correction_ru_regulated_3}
\Gamma_{r_{u}}(0) & = & \frac{2\lambda^{(d_{c}^{\lambda})} A_{v}^{(d_{c}^{\lambda},1)}r_{v}}{D_{v}} K_{d_{c}^{\lambda}}K_{1}\left(\frac{1}{\epsilon + 2\delta}\right).
\end{eqnarray}
We of course set $z=1$, corresponding to one temporal dimension. The
unusual $1/(\epsilon + 2\delta)$ pole in
Eq.~(\ref{eq:One_loop_correction_ru_regulated_3}) is a direct result
of the mixing between the frequency and momentum integrals, a
feature that does not appear in the purely spatial power-law noise
case.  It implies that the independent parameters $\epsilon$ and
$\delta$ do not have to be both zero to produce a divergence, only
the specific combination $\epsilon + 2\delta$ must vanish. This is
very different from the purely spatial power-law noise case. A
similar one-loop correction can be obtained for the rate constant
$\lambda$ (see bottom diagram in
Fig.~\ref{fig:One_loop_corrections}):
\begin{eqnarray}
\label{eq:One_loop_correction_lambda_regulated_3}
\Gamma_{\lambda}(0) & = & - 8\lambda_{(d_{c}^{\lambda})}^{2}A_{v}^{(d_{c}^{\lambda},1)} \frac{D_{u}}{D_{v}^{2}-D_{u}^{2}} \ln\left(\frac{D_{v}}{D_{u}}\right)   K_{d_{c}^{\lambda}} K_{1}  \left(\frac{1}{\epsilon}\right). \nonumber \\
\end{eqnarray}
Details of the computation are presented in
Appendix~\ref{sec:Correction_lambda}. Note that the interplay
between $\Gamma$-functions in
Eq.~(\ref{eq:One_loop_correction_ru_regulated_2}) may produce
unexpected cancellations of poles in the final result. For instance,
it is possible to show that  $\Gamma_{r_{u}}(0)$ has no pole and
thus finite for even $n$ and arbitrary $m$. This unexpected behavior
is absent from the purely spatially correlated power-law noise case.

\subsection{$\boldsymbol{\beta}$-function computations \label{sec:Beta_function}}

In the ($m=1$, $n=1$) case, the one-loop correction to the decay rate $\Gamma_{r_{u}}$ is divergent
when $d_{s} \geq y_{v}$ and the correction to the rate constant $\Gamma_{\lambda}$ is divergent for $d_{s} \geq y_{v} + 2$
(when $\epsilon = \delta = 0$).  For $\beta$-function computation purposes, there are thus three different regimes to distinguish.

\subsubsection{Regime 1} For $d_{s} < y_{v}$, both $r_{u}$ and $\lambda$ are finite and do not require renormalization.
The $\beta$-functions are trivial in this regime.

\subsubsection{Regime 2} For $y_{v} \leq d_{s} < y_{v} + 2$, $r_{u}$
is logarithmically divergent and $\lambda$ is finite.
Thus only $r_{u}$ requires renormalization and has a nontrivial $\beta$-function.
To study such a regime, we expand around the critical dimension $d_{c}^{\lambda}$ such
that $d = d_{c}^{\lambda} - (\epsilon + 2\delta)$ (with $\epsilon + 2\delta > 0$).
Following standard procedures, we write down the $Z$-factor for $r_{u}$:
\begin{eqnarray}
\label{eq:Z_factor_ru}
Z_{r_{u}} & = & 1 + \frac{\Gamma_{r_{u}}(0)}{r_{u}}, \\
          & = & 1 + 2g^{(d_{c}^{\lambda},1)} K_{d}K_{1}  \left(\frac{1}{\epsilon + 2\delta}\right),
\end{eqnarray}
where the effective coupling $g^{(d,z)}$ is defined as:
\begin{eqnarray}
\label{eq:Effective_coupling_g}
g^{(d,z)} & = & \frac{\lambda^{(d)}A_{v}^{(d,z)}r_{v}}{D_{v}r_{u}}.
\end{eqnarray}
Note that $g^{(d,z)}$ is dimensionless when $d\rightarrow
d_{c}^{\lambda}$ and $z\rightarrow 1$, as it should be. The
$\beta$-function for $g^{(d_{c}^{\lambda},1)}$ can be obtained from
Eq.~(\ref{eq:Effective_coupling_g}):
\begin{eqnarray}
\label{eq:Beta_function_g}
\beta_{g} & \equiv & T\frac{dg}{dT} \;\;=\;\; \frac{1}{2}\left|\epsilon + 2\delta\right|g + g^{2}K_{d}K_{1},
\end{eqnarray}
where $T$ is an arbitrary sliding scale and we dropped the $(d_{c}^{\lambda},1)$ superscript
to avoid cluttering of indices.  To interpret physically, we revert back to the original parameters with a
change of variable in Eq.~(\ref{eq:Beta_function_g}), giving:
\begin{eqnarray}
\label{eq:Beta_function_ru_regime2}
\beta_{r_{u}} & = & -\left(\frac{1}{2}|\epsilon + 2\delta|r_{u} + \frac{\lambda A_{v}r_{v}}{D_{v}} K_{d}K_{1}\right).
\end{eqnarray}
Perturbation theory limits the validity of the above result to the region where $g = \lambda A_{v}r_{v}/D_{v}r_{u} < 1$.

\subsubsection{Regime 3} For $d_{s} \geq y_{v} + 2$, both $r_{u}$ and $\lambda$ and
require renormalization.  Both have a nontrivial $\beta$-function.
To study this regime we expand around $d_{c}^{\lambda}$ such that $d
= d_{c}^{\lambda} - (\epsilon + 2\delta)$ (with $\epsilon + 2\delta
< 0$).  Since both $r_{u}$ and $\lambda$ run simultaneously, there
is an extra $Z$-factor in addition to Eq.~(\ref{eq:Z_factor_ru}):
\begin{eqnarray}
\label{eq:Z_factor_lambda}
Z_{\lambda} & = & 1 + \frac{\Gamma_{\lambda}(0)}{\lambda}, \\
            & = & 1 - 8h^{(d_{c}^{\lambda},1)}\ln\left(\frac{D_{v}}{D_{u}}\right)K_{d_{c}^{\lambda}}K_{1}\left(\frac{1}{\epsilon}\right),
\end{eqnarray}
where the effective coupling $h^{(d,z)}$ is defined as:
\begin{eqnarray}
\label{eq:Effective_coupling_h}
h^{(d,z)} & = & \frac{\lambda^{(d)}A_{v}^{(d,z)}D_{u}}{(D_{v}^{2}-D_{u}^{2})}.
\end{eqnarray}
Using Eqs.~(\ref{eq:Effective_coupling_g}) and (\ref{eq:Effective_coupling_h}),
the $\beta$-functions for the effective couplings $g^{(d_{c}^{\lambda},1)}$ and $h^{(d_{c}^{\lambda},1)}$ can be obtained:
\begin{eqnarray}
\label{eq:Beta_function_g_regime3}
\beta_{g} \;\;\equiv\;\; T\frac{dg}{dT} & = & \left(\frac{\epsilon}{2} + \delta\right)g + g^{2}K_{d}K_{1} \nonumber \\
                                        &   & + 8gh\ln\left(\frac{D_{v}}{D_{u}}\right)K_{d}K_{1}\left(\frac{1}{2} + \frac{\delta}{\epsilon}\right), \\
\label{eq:Beta_function_h_regime3}
\beta_{h} \;\;\equiv\;\; T\frac{dh}{dT} & = & \left(\frac{\epsilon}{2} + \delta\right)h \nonumber \\
                                        &   & + 8h^{2}\ln\left(\frac{D_{v}}{D_{u}}\right)K_{d}K_{1} \left(\frac{1}{2} + \frac{\delta}{\epsilon}\right).
\end{eqnarray}
Changing variables, we finally get the $\beta$-functions for the decay rate and rate constant in regime 3 in terms of the original parameters:
\begin{eqnarray}
\label{eq:Beta_function_ru_regime3}
\beta_{r_{u}} & = & -\frac{\lambda A_{v}r_{v}}{D_{v}} K_{d}K_{1}, \\
\label{eq:Beta_function_lambda_regime3}
\beta_{\lambda} & = & \left(\frac{\epsilon}{2} + \delta\right)\lambda \nonumber \\
                &   & + 8\lambda^{2}\frac{A_{v}D_{u}}{(D_{v}^{2}-D_{u}^{2})}\ln\left(\frac{D_{v}}{D_{u}}\right)K_{d}K_{1} \left(\frac{1}{2} + \frac{\delta}{\epsilon}\right),
\end{eqnarray}
with $\epsilon + 2\delta < 0$.
Perturbation theory limits the validity of the above results to the region where $g = \lambda A_{v}r_{v}/D_{v}r_{u}<1$
and $h = \lambda A_{v}D_{u}/(D_{v}^{2} - D_{u}^{2}) <1$.  Note that there seems to be a potential
divergence in $\beta_{\lambda}$ when $\delta \neq 0$ and $\epsilon \rightarrow 0$.  This is problematic, and might signal
a missing contribution in the perturbative expansion.  Fortunately, this divergence is only apparent and can be understood in the following way.
The corrections~(\ref{eq:One_loop_correction_ru_regulated_3}) and~(\ref{eq:One_loop_correction_lambda_regulated_3})
diverge when $\epsilon + 2\delta \rightarrow 0$ and $\epsilon\rightarrow 0$, respectively.  Both corrections diverge in
regime 3, implying that $\epsilon$ and $\delta$ must tend to zero simultaneously in this regime.  Consequently, the situation
$\delta \neq 0$ and $\epsilon\rightarrow 0$ does not correspond to regime 3, and thus cannot be included in the analysis of Eq.~(\ref{eq:Beta_function_lambda_regime3}).

\section{Results and discussion \label{sec:Discussion}}

In this section we integrate the $\beta$-functions obtained in Sect.~\ref{sec:Renormalization} in
order to study the behavior of parameters at smaller scales.  We also compare the running solutions
to the purely spatial power-law noise case, and point out qualitative differences in behavior.
The $\beta$-functions in regime 1 are trivial, thus we only consider regimes 2 and 3 in the following.

\subsection{Running of parameters in regime 2 \label{sec:Running_parameters_regime2}}

\begin{figure}[H]
\includegraphics[width=0.49\textwidth]{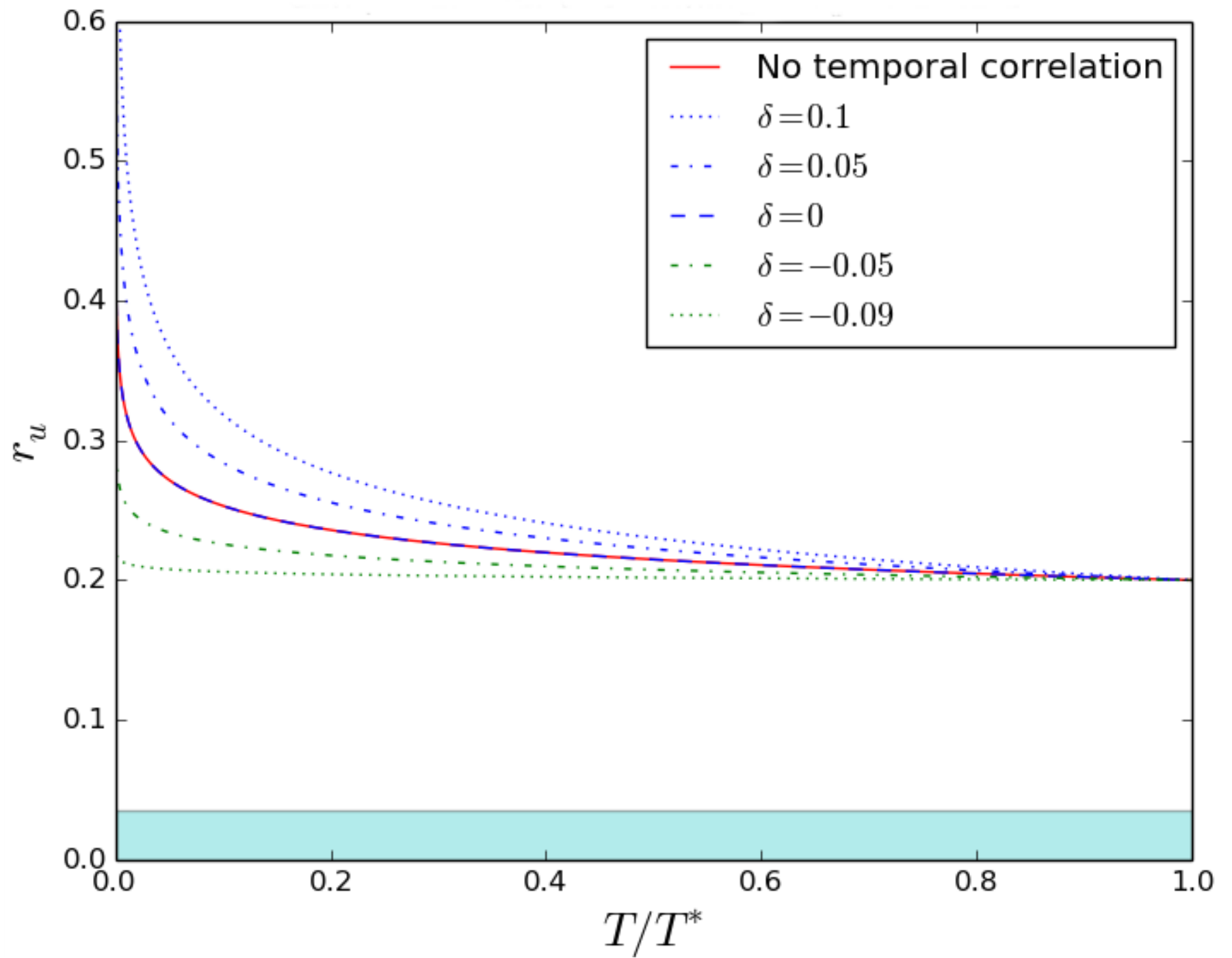}%
\caption{Running of $r_{u}$ for regime 2 for different values of
$\delta$ after effecting the rescaling in
Eq.~(\ref{eq:Rescaling_regime2}). The red line corresponds to the
purely spatial correlation case (c.f.
Eq.~(\ref{eq:Running_ru_regime2_GHP})). We used $\epsilon = 0.2$,
$D_{v} = 0.3$, $r_{v} = 0.4$, $A_{v}^{\rm (GHP)} = 0.1$, $\lambda =
0.05$,  $K_{d} = 0.05$, $r_{u}(T^{*}) = 0.2$ for the plotting. The
shaded region indicates the breakdown of perturbation theory.
\label{fig:Running_ru_regime2}}
\end{figure}

The running of the decay rate in regime 2 is obtained by integrating the $\beta$-function~(\ref{eq:Beta_function_ru_regime2}).  The result is:
\begin{eqnarray}
\label{eq:Running_ru_regime2}
r_{u}(T) & = & \left(r_{u}(T^{*}) + \frac{2\lambda A_{v}r_{v}}{|\epsilon + 2\delta|D_{v}}K_{d}K_{1}\right)\left(\frac{T}{T^{*}}\right)^{-\frac{|\epsilon + 2\delta|}{2}} \nonumber \\
         &   & - \frac{2\lambda A_{v}r_{v}}{|\epsilon + 2\delta|D_{v}}K_{d}K_{1},
\end{eqnarray}
where $T^{*}$ is some timescale at which $r_{u}(T^{*})$ is known and can be measured.
The result~(\ref{eq:Running_ru_regime2}) can be compared to the purely spatial power-law noise case
found in Ref.~\cite{Gagnon_etal_2015} (denoted by a GHP superscript in what follows):
\begin{eqnarray}
\label{eq:Running_ru_regime2_GHP}
r_{u}^{\rm (GHP)}(T) &  = & \left(r_{u}(T^{*}) + \frac{\lambda A_{v}^{\rm (GHP)}r_{v}K_{d}}{|\epsilon|D_{v}^{2}}\right)\left(\frac{T}{T^{*}}\right)^{-|\epsilon|/2} \nonumber \\
                     &    & - \frac{\lambda A_{v}^{\rm (GHP)}r_{v} K_{d}}{|\epsilon|D_{v}^{2}}.
\end{eqnarray}
We note that Eqs.~(\ref{eq:Running_ru_regime2}) and
(\ref{eq:Running_ru_regime2_GHP}) are very similar, but differ in
some aspects.  First, there are extra numerical factors appearing in
Eq.~(\ref{eq:Running_ru_regime2}) due to the nontrivial integral
over frequency.  There is also a factor of $D_{v}$ absent from
Eq.~(\ref{eq:Running_ru_regime2}) due to the fact that the
engineering dimensions of the noise amplitudes $A_{v}$ and
$A_{v}^{\rm (GHP)}$ are different in both cases.  Those changes are
quantitative in nature and do not produce any qualitative changes in
the running of the decay rate.  They can be eliminated by a simple
rescaling of the noise amplitude in the temporally correlated case.
Substituting the noise amplitude $A_{v}$ in
Eq.~(\ref{eq:Running_ru_regime2}) with:
\begin{eqnarray}
\label{eq:Rescaling_regime2}
A_{v} & \rightarrow & \frac{A_{v}^{\rm (GHP)}}{2K_{1}D_{v}},
\end{eqnarray}
where $A_{v}^{\rm (GHP)}$ is the noise amplitude in the purely spatial power-law noise case, we obtain:
\begin{eqnarray}
\label{eq:eq:Running_ru_regime2_rescaled}
r_{u}(T) & = & \left(r_{u}(T^{*}) + \frac{\lambda A_{v}^{\rm (GHP)}r_{v}}{|\epsilon + 2\delta|D_{v}^{2}}K_{d}\right)\left(\frac{T}{T^{*}}\right)^{-\frac{|\epsilon + \delta|}{2}} \nonumber \\
         &   & - \frac{\lambda A_{v}^{\rm (GHP)}r_{v}}{|\epsilon + 2\delta|D_{v}^{2}}K_{d}.
\end{eqnarray}
At the pole (i.e. when $\delta = 0$), there is an exact mapping between the purely
spatial case~(\ref{eq:Running_ru_regime2_GHP}) and the one with a mixture of spatial and
temporal noise correlations~(\ref{eq:eq:Running_ru_regime2_rescaled}).  Away from the pole (i.e. when $\delta \neq 0$), the
behaviors are different, as shown in Fig.~\ref{fig:Running_ru_regime2}.

In the living system interpretation of the CARD model, a change in
decay rate $r_u$ due to fluctuations implies a change at which
nutrient $U$ is removed from the system. Since the growth of an
organism (i.e. $\partial V/\partial t$) is proportional to the
amount of food present (i.e. $\lambda UV^{2}$), a larger decay rate
implies smaller growth of structures (and vice-versa). From
Fig.~\ref{fig:Running_ru_regime2}, the value of the decay rate is
greater or smaller (with respect to the purely spatial case) at
shorter scales, depending on the value of $\delta$.  This means that
temporal correlations can either increase or decrease the running of
$r_{u}$, and thus enhance or suppress the growth of structures at
small temporal scales.

%

\subsection{Running of parameters in regime 3 \label{sec:Running_parameters_regime3}}

The running solutions for the decay rate and rate constant in regime 3 are obtained by integrating
the $\beta$-functions~(\ref{eq:Beta_function_ru_regime3}) and~(\ref{eq:Beta_function_lambda_regime3}), giving:
\begin{widetext}
\begin{eqnarray}
\label{eq:Running_ru_regime3}
r_{u}(T) & = & r_{u}(T^{*}) - \frac{\epsilon}{|\epsilon + 2\delta|} \frac{r_{v}(D_{v}^{2}-D_{u}^{2})}{4D_{v}D_{u}\ln\left(\frac{D_{v}}{D_{u}}\right)} \ln\left[1 - \frac{8A_{v}D_{u} \ln\left(\frac{D_{v}}{D_{u}}\right) K_{d}K_{1} \lambda(T^{*})}{\epsilon (D_{v}^{2}-D_{u}^{2})} \left(\left(\frac{T}{T^{*}}\right)^{-\frac{|\epsilon + 2\delta|}{2}} -1 \right) \right], \\
\label{eq:Running_lambda_regime3}
\lambda(T) & = & -\frac{\epsilon (D_{v}^{2}-D_{u}^{2})}{8A_{v}D_{u}\ln\left(\frac{D_{v}}{D_{u}}\right)K_{d}K_{1}} \left(\frac{1}{1 - \left(1 + \frac{\epsilon (D_{v}^{2}-D_{u}^{2})}{8A_{v}D_{u}\ln\left(\frac{D_{v}}{D_{u}}\right)K_{d}K_{1} \lambda(T^{*})}\right)\left(\frac{T}{T^{*}}\right)^{\frac{|\epsilon + 2\delta|}{2}}}\right),
\end{eqnarray}
\end{widetext}
with the condition $\epsilon + 2\delta < 0$.  Here again $T^{*}$ is some timescale
at which $r_{u}(T^{*})$ and $\lambda(T^{*})$ are known.  The above running
solutions~(\ref{eq:Running_ru_regime3})-(\ref{eq:Running_lambda_regime3}) can be compared
to the purely spatial power-law noise results of Ref.~\cite{Gagnon_etal_2015}:
\begin{widetext}
\begin{eqnarray}
\label{eq:Running_ru_regime3_GHP}
r_{u}^{\rm (GHP)}(T) & = & r_{u}(T^{*}) + \frac{r_{v}(D_{u}+D_{v})}{4D_{v}}  \ln\left[1 + \frac{4A_{v}^{\rm (GHP)}K_{d}\lambda(T^{*})}{|\epsilon|D_{v}(D_{u}+D_{v})} \left(\left(\frac{T}{T^{*}}\right)^{-\frac{|\epsilon|}{2}} -1 \right) \right], \\
\label{eq:Running_lambda_regime3_GHP}
\lambda^{\rm (GHP)}(T) & = & \frac{|\epsilon|D_{v}(D_{u}+D_{v})}{4A_{v}^{\rm (GHP)}K_{d}}  \left(\frac{1}{1 - \left(1 - \frac{|\epsilon|D_{v}(D_{u}+D_{v})}{4A_{v}^{\rm (GHP)}K_{d}\lambda(T^{*})}\right) \left(\frac{T}{T^{*}}\right)^{\frac{|\epsilon|}{2}}}\right).
\end{eqnarray}
\end{widetext}
Just as for regime 2, the two sets of running solutions are very similar, but differ in some crucial places.  Many of those numerical and dimensional factor differences can be eliminated using the following rescaling of the noise amplitude in Eqs.~(\ref{eq:Running_ru_regime3})-(\ref{eq:Running_lambda_regime3}):
\begin{eqnarray}
\label{eq:Rescaling_regime3}
A_{v} & \rightarrow & \frac{A_{v}^{\rm (GHP)} (D_{v}-D_{u})}{2K_{1}D_{v}D_{u}\ln\left(\frac{D_{v}}{D_{u}}\right)}.
\end{eqnarray}
Doing the above substitution, we get:
\begin{widetext}
\begin{eqnarray}
\label{eq:Running_ru_regime3_rescaled}
r_{u}(T) & = & r_{u}(T^{*}) - \frac{\epsilon}{|\epsilon + 2\delta|} \frac{r_{v}(D_{v}+D_{u})}{4D_{v}}\left[\frac{(D_{v}-D_{u})}{D_{u}\ln\left(\frac{D_{v}}{D_{u}}\right)}\right]  \ln\left[1 - \frac{4A_{v}^{\rm (GHP)} K_{d} \lambda(T^{*})}{\epsilon D_{v}(D_{v}+D_{u})} \left(\left(\frac{T}{T^{*}}\right)^{-\frac{|\epsilon + 2\delta|}{2}} -1 \right) \right], \\
\label{eq:Running_lambda_regime3_rescaled}
\lambda(T) & = & -\frac{\epsilon D_{v}(D_{v}+D_{u})}{4A_{v}^{\rm (GHP)}K_{d}} \left(\frac{1}{1 - \left(1 + \frac{\epsilon D_{v}(D_{v}+D_{u})}{4A_{v}^{\rm (GHP)}K_{d} \lambda(T^{*})}\right)\left(\frac{T}{T^{*}}\right)^{\frac{|\epsilon + 2\delta|}{2}}}\right).
\end{eqnarray}
\end{widetext}
When $\delta = 0$ (and taking into account the condition $\epsilon+2\delta<0$), we see that Eq.~(\ref{eq:Running_lambda_regime3_GHP}) is identical to Eq.~(\ref{eq:Running_lambda_regime3_rescaled}), showing an exact mapping between the purely spatial and spatial+temporal mixture cases for the rate constant.  No such mapping exists for the decay rate because of the presence of extra diffusion constant factors that cannot be rescaled away (compare Eqs.~(\ref{eq:Running_ru_regime3_GHP}) and~(\ref{eq:Running_ru_regime3_rescaled})).

\begin{figure}
\parbox{0.49\textwidth}{\includegraphics[width=0.49\textwidth]{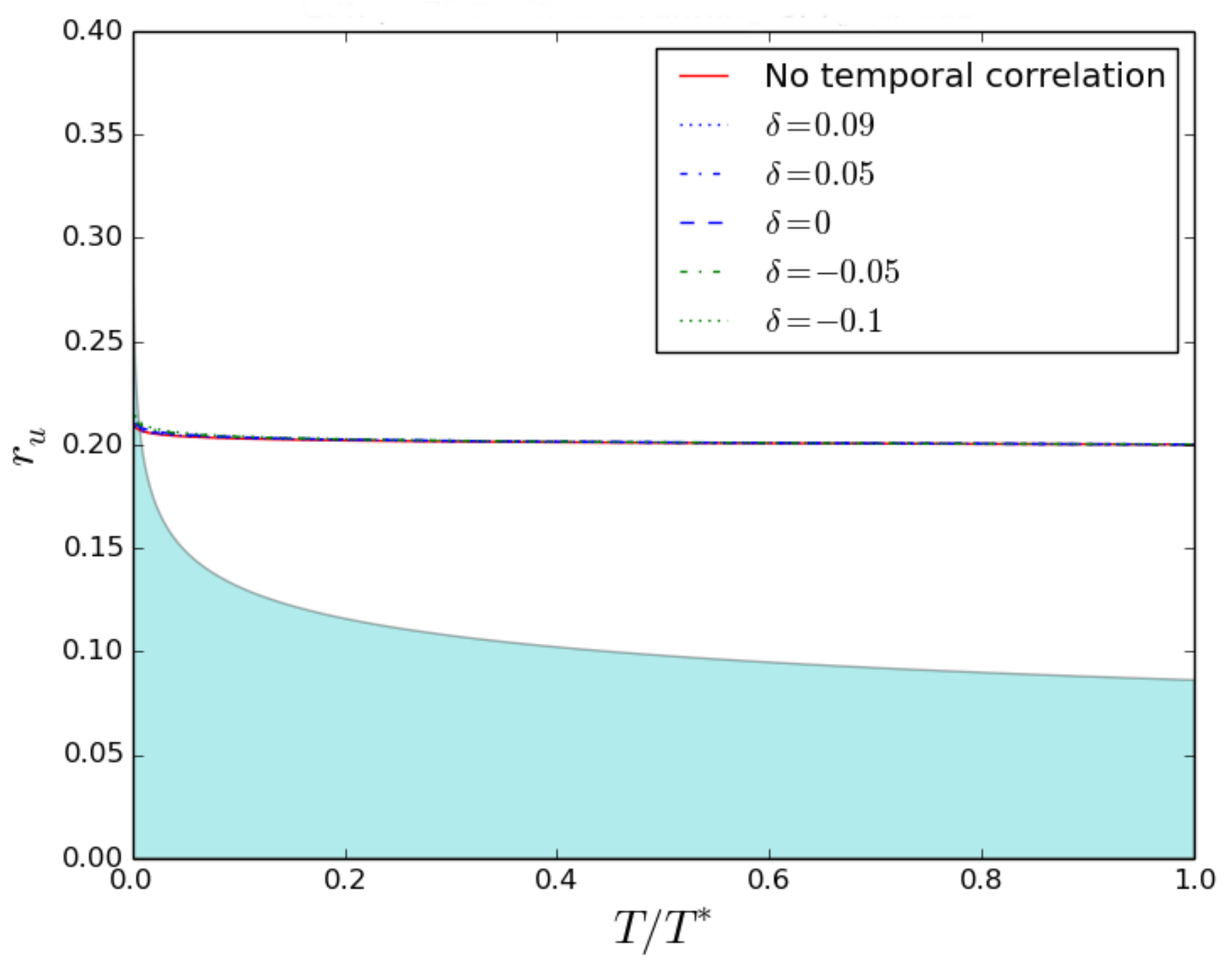}} \\
\parbox{0.49\textwidth}{\includegraphics[width=0.49\textwidth]{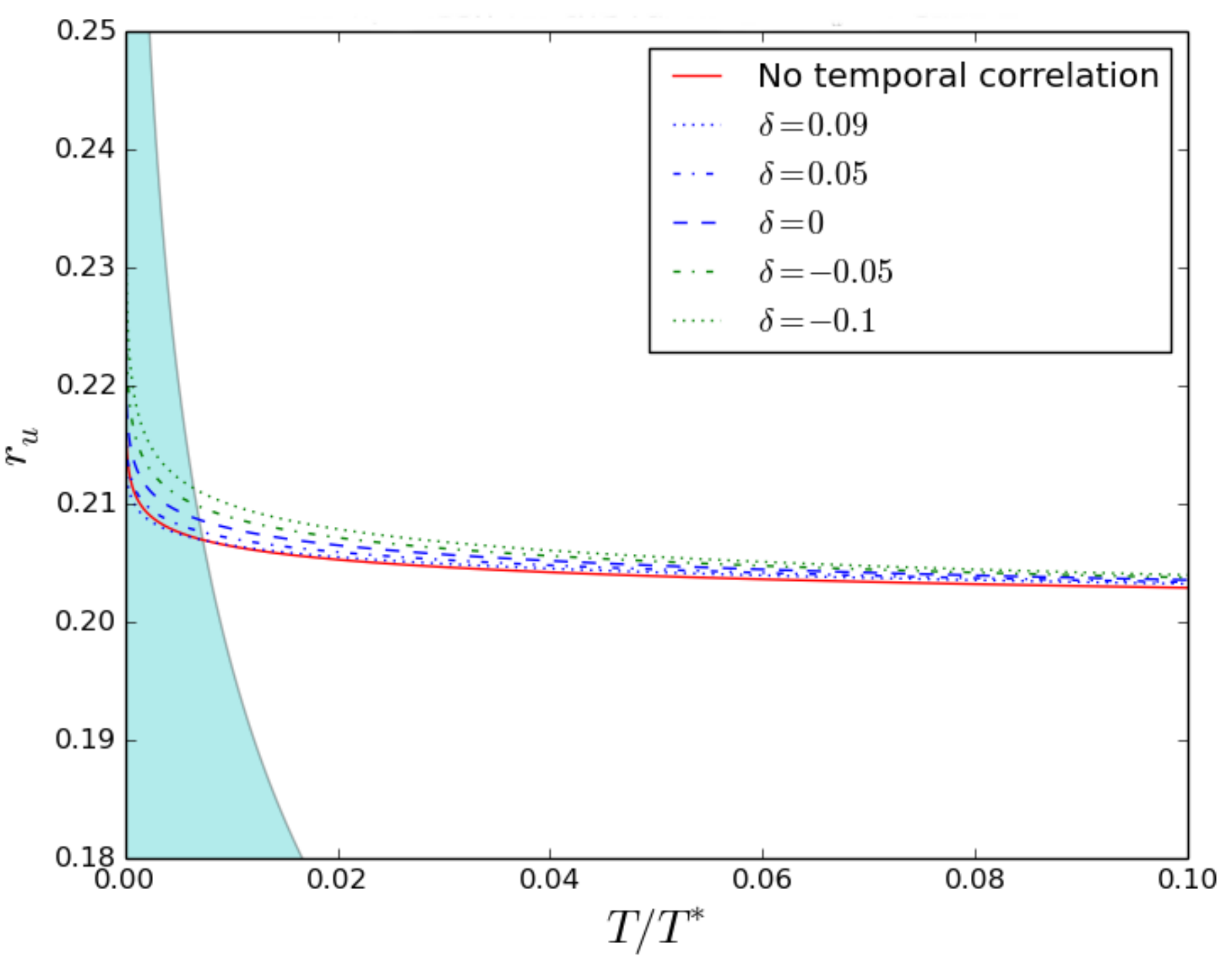}}
\caption{Top: Running of $r_{u}$ for regime 3 for different values of $\delta$ after doing the rescaling in Eq.~(\ref{eq:Rescaling_regime3}).  The red line corresponds to the purely spatial correlation case (c.f. Eq.~(\ref{eq:Running_ru_regime3_GHP})).  We used $\epsilon = -0.2$, $D_{v} = 0.3$, $D_{u} = 0.2{\tiny }$, $r_{v} = 0.4$, $A_{v}^{\rm (GHP)} = 0.1$, $K_{d} = 0.05$, $r_{u}(T^{*}) = 0.2$, $\lambda(T^{*}) = 0.1$ for the plotting.  The shaded region indicates the breakdown of perturbation theory.  Bottom: Zoom of the top figure.
\label{fig:Running_ru_regime3}}
\end{figure}

Effects of a nonzero $\delta$ on the running of the decay rate are small, since the power-law term $\left(T/T^{*}\right)^{-\frac{|\epsilon + 2\delta|}{2}}$ is inside a logarithm.  This can be seen in Fig.~\ref{fig:Running_ru_regime3}.  The region where deviations are the largest are in the shaded area where perturbation theory cannot be trusted.  Thus we conclude that power-law temporal correlations have a negligible effect on the decay rate in regime 3.

\begin{figure}
\includegraphics[width=0.49\textwidth]{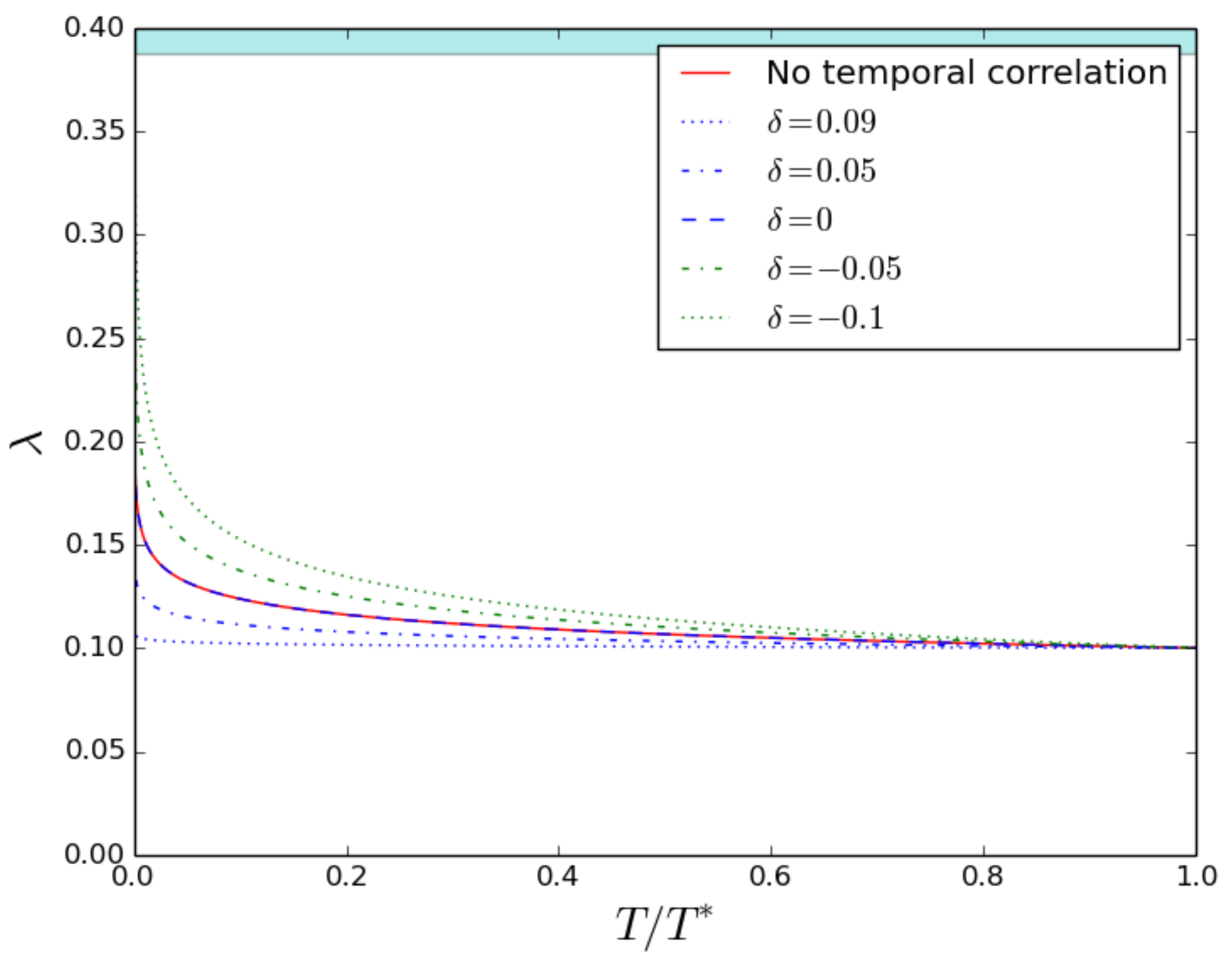}
\caption{Running of $\lambda$ for regime 3 for different values of $\delta$ after doing the rescaling in Eq.~(\ref{eq:Rescaling_regime3}).  The red line corresponds to the purely spatial correlation case (c.f. Eq.~(\ref{eq:Running_lambda_regime3_GHP})).  We used $\epsilon = -0.2$, $D_{v} = 0.3$, $D_{u} = 0.2{\tiny }$, $A_{v}^{\rm (GHP)} = 0.1$,  $K_{d} = 0.05$, $\lambda(T^{*}) = 0.1$ for the plotting.  The shaded region indicates the breakdown of perturbation theory.
\label{fig:Running_lambda_regime3}}
\end{figure}

The effect of a nonzero $\delta$ on the catalysis rate constant is
more pronounced than for the $U$-decay rate. This can be seen on
Fig.~\ref{fig:Running_lambda_regime3}.  In the living system
interpretation of the CARD model, a change in constant rate due to
fluctuations implies a change in the rate at which new ``body
parts'' are created. Since the growth of an organism (i.e. $\partial
V/\partial t$) is proportional to $\lambda UV^{2}$, an increase in
$\lambda$ leads to a larger growth of structures (and vice-versa).
From Fig.~\ref{fig:Running_lambda_regime3}, we see that the value of
the rate constant is larger or smaller at shorter scales with
respect to the purely spatial case, depending on the distance from
the pole $\delta$. Thus nontrivial temporal correlations can enhance
or suppress the growth of structures at small scales.

\section{Conclusion \label{sec:Conclusion}}

In general, large and small scale environmental fluctuations can have nontrivial effects on the
small-scale, kinetic and collective properties of chemical systems. In this paper, we have
analyzed the effect of noise with both spatial and temporal power-law
correlations on the kinetics and phenomenology of a specific cubic autocatalytic reaction-diffusion
chemical system.  The noise is additive, in that its presence is incorporated as an additive term affecting the individual reaction rates.  In particular, using renormalization group 
techniques, we show analytically how parameters such as chemical decay rates and catalytic 
rate constants depend on the statistical properties of the noise. The noise
causes the model parameters to \textit{renormalize}, that is, to change due to the simultaneous  
presence of fluctuations and nonlinearities. These noise-induced changes 
manifest themselves as an inherited and explicit scale dependence of the parameters so
affected. Some of these renormalization effects have been studied previously for the case of purely spatial noise in Ref.~\cite{Gagnon_etal_2015}.  Here we found that those effects can be greatly enhanced or suppressed with the presence of temporally correlated noise.  We also discuss how under certain conditions, the effects of spatial and temporal noises can be mapped onto each other.

These results raise important questions regarding the role of external noise in both chemical and biological self-organization, and in the environmental selection of reproduction (catalysis) as well as other dynamical mechanisms. We learn that when describing an open chemical system, not only must we estimate the relevant parameters, but also the magnitude of the stochastic influences, the spatial and temporal scales, and the correlations of the latter.  The interdependence between an open chemical system and its environment can therefore have important consequences. In the case of nonlinear systems, of which chemical reactions provide immediate examples, the role of noise can be nontrivial, by forcing and driving the system to explore new situations which although not present in the purely deterministic situation might be favored by external natural selection or enhancement processes and their subsequent pressures. For an adaptive system this can have a direct impact on how the system evolves.

Understanding the effect of external noise on chemical systems can have many applications. One class of applications considers the noise as environmental.  Since the reaction-diffusion model studied in this paper can be viewed as a (very simple) generic prototype model for a living system, the techniques developed in the present paper could be used to study its viability\cite{Egbert_PerezMercader_2016} under time-dependent environmental pressures.  Another class of applications considers the noise as an experimental tool or probe to uncover underlying mechanisms and pathways in chemical systems~\cite{Gagnon_PerezMercader_2015}.  The work presented in this paper extends the range of possible noise correlations that can be used to study chemical systems experimentally.  The application of the present results to the above problems is a topic of our current research and study.

\appendix

\section{Feynman rules \label{sec:Feynman_rules}}

A general discussion of Feynman rules for stochastic partial
differential equations can be found in
Ref.~\cite{Barabasi_Stanley_1995}. In the present case, they are
obtained by iterating the Fourier-transformed stochastic CARD
equations~(\ref{eq:Gray_Scott_equations_1})-(\ref{eq:Gray_Scott_equations_2})
and identifying each component with a picture.  Free response
functions are given by (see Fig.~\ref{fig:Feynman_rules}):
\begin{eqnarray}
G_{v0}(k) & = & \frac{1}{D_{v}|{\bf k}|^{2} - i\omega + r_{v}}, \\
G_{u0}(k) & = & \frac{1}{D_{u}|{\bf k}|^{2} - i\omega + r_{u}}.
\end{eqnarray}
with arrows following the sign of the frequency.  Tree-level interactions are given by (see Fig.~\ref{fig:Feynman_rules}):
\begin{figure}
\includegraphics[width=0.45\textwidth]{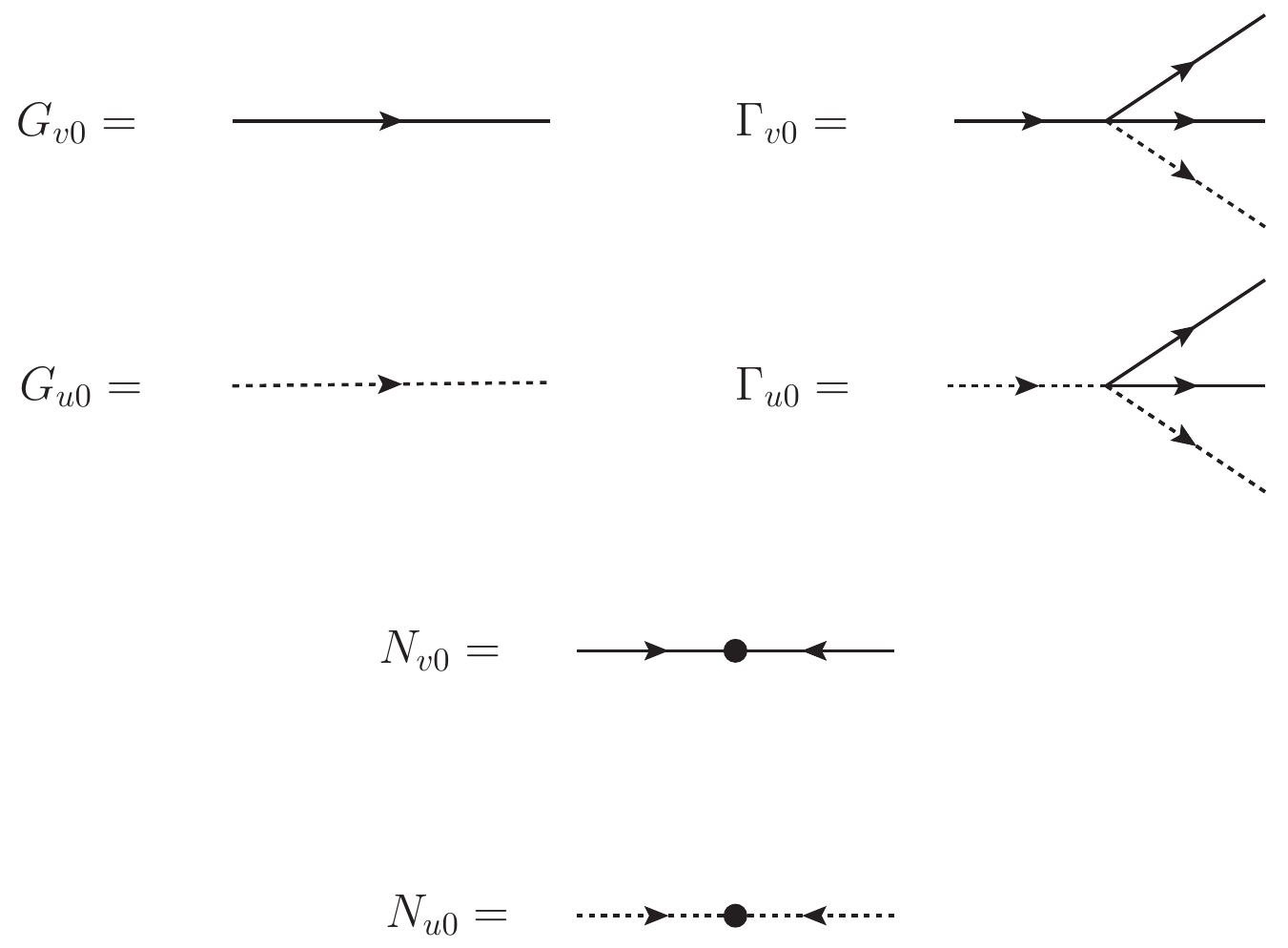}%
\caption{Feynman rules for the stochastic CARD model corresponding to Eqs.~(\ref{eq:Gray_Scott_equations_1})-(\ref{eq:Gray_Scott_equations_2}).
\label{fig:Feynman_rules}}
\end{figure}
\begin{eqnarray}
\Gamma_{v0} & = & -\Gamma_{u0} \;\;=\;\; \lambda.
\end{eqnarray}
The effect of fluctuations is represented by loop diagrams in field theory.  Loop diagrams are obtained from noise averaging (c.f. Eqs.~(\ref{eq:Noise_property_2})-(\ref{eq:Noise_property_3})):
\begin{eqnarray}
N_{v0}(k) & = & 2A_{v}|{\bf k}|^{-y_{v}}\omega^{-2\theta_v}, \\
N_{u0}(k) & = & 2A_{u}|{\bf k}|^{-y_{u}}\omega^{-2\theta_u},
\end{eqnarray}
The components shown in Fig.~\ref{fig:Feynman_rules}, supplemented with conservation of momentum at each vertex and integration over undetermined momenta, form the basis of perturbation theory.  With the appropriate combinatoric factor, they can be used to write down any Feynman diagram for the stochastic CARD model.

\vspace{\baselineskip}

\section{Computation of the one-loop correction to the rate constant $\lambda$ \label{sec:Correction_lambda}}

The expression for the one-loop correction to the rate constant is:
\begin{widetext}
\begin{eqnarray}
\label{eq:Integration_vertex_function_1}
\Gamma_{\lambda}(0) & = & -8\lambda^{2}A_{v} \int\frac{d^{d_{s}}p}{(2\pi)^{d_{s}}}\int\frac{d\omega}{(2\pi)}\; \omega^{-2\theta_{v}} |{\bf p}|^{-y_{v}} \left(\frac{1}{D_{u}|{\bf p}|^{2} + i\omega + r_{u}}\right) \left(\frac{1}{D_{v}|{\bf p}|^{2} - i\omega + r_{v}}\right) \left(\frac{1}{D_{v}|{\bf p}|^{2} + i\omega + r_{v}}\right), \nonumber \\
                    & = & -8\lambda^{2}A_{v} \int\frac{d^{d_{s}}p}{(2\pi)^{d_{s}}}
\int\frac{d\omega}{(2\pi)}\; |\omega|^{-2\theta_{v}} |{\bf p}|^{-y_{v}}\; \left(\frac{1}{|\omega|^{2} + (D_{v}|{\bf p}|^{2} + r_{v})^{2}}\right) \left(\frac{D_{u}|{\bf p}|^{2} + r_{u}}{|\omega|^{2} + (D_{u}|{\bf p}|^{2} + r_{u})^{2}}\right).
\end{eqnarray}
\end{widetext}
Depending on the parameters $d_{s}$, $y_{v}$ and $\theta_{v}$, the above one-loop correction may diverge.  We regulate the expression by analytically continuing the space dimension to $d$ and the time dimension to $z$, giving:
\begin{widetext}
\begin{eqnarray}
\label{eq:Integration_vertex_function_2}
\Gamma_{\lambda}(0) & = & -8\lambda_{(d)}^{2}A_{v}^{(d,z)} \int\frac{d^{d}p}{(2\pi)^{d}}
\int\frac{d^{z}\omega}{(2\pi)^{z}}\; |\omega|^{-2\theta_{v}} |{\bf p}|^{-y_{v}}\; \left(\frac{1}{|\omega|^{2} + (D_{v}|{\bf p}|^{2} + r_{v})^{2}}\right) \left(\frac{D_{u}|{\bf p}|^{2} + r_{u}}{|\omega|^{2} + (D_{u}|{\bf p}|^{2} + r_{u})^{2}}\right).
\end{eqnarray}
\end{widetext}
Simple power counting shows that the one-loop correction to the rate constant behaves as:
\begin{eqnarray}
\label{eq:Power_counting_lambda}
\Gamma_{\lambda} & \sim & \Lambda^{d-y_{v}+2z-4\theta_{v}-6} \sim \Lambda^{2m + 2n - 4 - \epsilon -2\delta}.
\end{eqnarray}
From Eq.~(\ref{eq:Power_counting_lambda}), we infer that $\Gamma_{\lambda}$ is logarithmically UV divergent for ($m=2$, $n=0$) and ($m = 1$, $n = 1$).  The first case corresponds to purely spatial noise and is treated in Ref.~\cite{Gagnon_etal_2015}.  The second case is a mixture of spatial and temporal noise.  We focus on the latter in the following.

We start by doing the integral over frequency first.  Introducing a Feynman parameter in Eq.~(\ref{eq:Integration_vertex_function_2}), we obtain:
\begin{eqnarray}
\label{eq:Integration_vertex_function_3}
\Gamma_{\lambda}(0) & = & -8\lambda_{(d)}^{2}A_{v}^{(d,z)} \int\frac{d^{d}p}{(2\pi)^{d}}\; |{\bf p}|^{-y_{v}} (D_{u}|{\bf p}|^{2} + r_{u}) \nonumber \\
                    &   & \int_{0}^{1}dx\int\frac{d^{z}\omega}{(2\pi)^{z}} |\omega|^{-2\theta_{v}} \left(\frac{1}{|\omega|^{2} + \Delta(x)}\right)^{2},
\end{eqnarray}
where
\begin{equation}
\Delta(x) = x(D_{v}|{\bf p}|^{2} + r_{v})^{2} + (1-x)(D_{u}|{\bf p}|^{2} + r_{u})^{2}.
\end{equation}
Integrating Eq.~(\ref{eq:Integration_vertex_function_3}) using the method of dimensional regularization in the presence of noise~\cite{Gagnon_etal_2015}, we obtain:
\begin{eqnarray}
\label{eq:Integration_vertex_function_4}
\lefteqn{\Gamma_{\lambda}(0)  =  -4\lambda_{(d)}^{2}A_{v}^{(d,z)} K_{z} \Gamma\left(-\frac{z}{2} +\theta_{v} + 2\right) \Gamma\left(\frac{z}{2}-\theta_{v}\right)} \nonumber \\
                    &   & \int\frac{d^{d}p}{(2\pi)^{d}} |{\bf p}|^{-y_{v}} (D_{u}|{\bf p}|^{2} + r_{u}) \int_{0}^{1}dx \frac{(\Delta(x))^{\frac{z}{2} -\theta_{v} +2}}{2}. \nonumber \\
\end{eqnarray}
The integral over the Feynman parameter can be done explicitly.  The result is:
\begin{widetext}
\begin{equation}
\label{eq:Integration_vertex_function_5}
\Gamma_{\lambda}(0) = 4\lambda_{(d)}^{2}A_{v}^{(d,z)} K_{z} \frac{\pi}{\sin\pi\left(\frac{z}{2}-\theta_{v}\right)} \int\frac{d^{d}p}{(2\pi)^{d}}\; |{\bf p}|^{-y_{v}} \frac{(D_{u}|{\bf p}|^{2} + r_{u}) \left[(D_{v}|{\bf p}|^{2} + r_{v})^{z-2\theta_{v} -2} - (D_{u}|{\bf p}|^{2} + r_{u})^{z-2\theta_{v} -2} \right]}{\left[(D_{v}|{\bf p}|^{2} + r_{v})^{2} - (D_{u}|{\bf p}|^{2} + r_{u})^{2}\right]}.
\end{equation}
\end{widetext}
To do the integration over momentum, another Feynman parameter must be introduced.  Before proceeding, we point out an important subtlety not present in usual quantum field theory computations.  The temporal noise correlation exponent $\theta_{v}$ is a parameter and is not specified in Eq.~(\ref{eq:Integration_vertex_function_5}).  This implies that the number of factors in the denominator in the integrand and their power depend on $\theta_{v}$.  In practice, the Feynman trick is used to regroup factors in the denominator only.  Since $\theta_{v}$ is left unspecified, the number of Feynman parameters needed to regroup the factors in the denominator is ambiguous.  This problem is not present in usual quantum field theory computation, since the number of factors is fixed.

To make progress, it is necessary to specify $\theta_{v}$.  As explained in Sect.~\ref{sec:One_loop_corrections}, the first nontrivial case of spatial and temporal noise mixing is ($m=1$, $n=1$), implying $z - 2\theta_{v} = 2 - \delta$.  Plugging this in Eq.~(\ref{eq:Integration_vertex_function_5}) and expanding for small $\delta$, we obtain:
\begin{widetext}
\begin{eqnarray}
\label{eq:Integration_vertex_function_6}
\Gamma_{\lambda}(0) & = & -8\lambda_{(d)}^{2}A_{v}^{(d,z)} \frac{D_{u}}{D_{v}^{2}-D_{u}^{2}}  K_{z} K_{d} \int_{0}^{\infty}d|{\bf p}|\; |{\bf p}|^{d-y_{v}-1} \frac{\left(|{\bf p}|^{2} + \frac{r_{u}}{D_{u}}\right) \left(\ln\left[\frac{D_{v}}{D_{u}}\right] + \ln\left[\frac{(|{\bf p}|^{2} + d_{1})}{(|{\bf p}|^{2} + d_{2})}\right]\right)}{\left(|{\bf p}|^{2} + d_{3}\right) \left(|{\bf p}|^{2} + d_{4}\right)},
\end{eqnarray}
\end{widetext}
where $d_{1} = \frac{r_{v}}{D_{v}}$, $d_{2} = \frac{r_{u}}{D_{u}}$, $d_{3} = \frac{r_{v}+r_{u}}{D_{v}+D_{u}}$ and $d_{4} = \frac{r_{v}-r_{u}}{D_{v}-D_{u}}$.  There are four contributions to the integrand.  Using a cutoff, a tedious calculation shows that only the contribution proportional to $|{\bf p}|^{2}\ln\left[\frac{D_{v}}{D_{u}}\right]$ leads to a logarithmic divergence, with all the other terms finite in the UV.  Discarding the finite terms, we introduce a Feynman parameter in the remaining term:
\begin{eqnarray}
\label{eq:Integration_vertex_function_7}
\Gamma_{\lambda}(0) & = & -8\lambda_{(d)}^{2}A_{v}^{(d,z)} \frac{D_{u}}{D_{v}^{2}-D_{u}^{2}} \ln\left(\frac{D_{v}}{D_{u}}\right)K_{z} K_{d} \nonumber \\
                    &   &  \int_{0}^{1}dx \int_{0}^{\infty}d|{\bf p}|\; |{\bf p}|^{d-y_{v}+1} \left(\frac{1}{|{\bf p}|^{2} + \zeta(x)}\right)^{2},
\end{eqnarray}
where
\begin{eqnarray}
\zeta(x) & = & xd_{3} + (1-x)d_{4}.
\end{eqnarray}
Integrating Eq.~(\ref{eq:Integration_vertex_function_7}) using the method of dimensional regularization in the presence of noise~\cite{Gagnon_etal_2015} and expanding around the critical dimension $d_{c}^{\lambda} = y_{v} + 2$, we finally obtain the one-loop correction to the rate constant $\lambda$ shown in Eq.~(\ref{eq:One_loop_correction_lambda_regulated_3}).

\begin{acknowledgments}
J.-S.G. and J.P.-M. thank Repsol S.A. for their support. D.H.
acknowledges Grant No. CTQ2013-47401-C2-2-P from MINECO (Spain).
\end{acknowledgments}

\bibliography{temporal_noise_GS}

\end{document}